\documentclass[twocolumn,aps,amsmath]{revtex4}
\usepackage{graphicx}
\usepackage{amssymb}
\usepackage{bm}

\newcommand\bnabla{\boldsymbol \nabla}
\newcommand\pll{ \parallel }

\begin{document}

\title{Interchange transport in electron-positron plasmas with ion impurities} 

\author{Alexander Kendl}
\affiliation{Institut f\"ur Ionenphysik und Angewandte Physik, Universit\"at
  Innsbruck, Technikerstr.~25, 6020 Innsbruck, Austria} 

\begin{abstract}
Interchange drive and cross-field transport of density filaments in
quasi-neutral inhomogeneously magnetized electron-positron plasmas is shown to
be strongly reduced by the presence of minority ions.
Two mechanisms are identified for the reduction in radial propagation and
plasma transport: effective mass related inertia, and collisionality dependent
Boltzmann spin-up of the filaments.  Numerical results are obtained with a
three-dimensional full-F multi-species gyrofluid model. 
\end{abstract} 

\maketitle

\section{Introduction}

Plans and first progress in laboratory confinement of quasi-neutral
electron-positron (e-p) plasmas in toroidal magnetic fields
\cite{danielson15,greaves97,tsytovich78,pedersen03,pedersen12}
has also generated renewed theoretical interest in magnetized e-p plasmas
\cite{helander14,helander16,zocco17,mishchenko18,mishchenko18B}.

It was recently shown that magnetic field-aligned density filaments (in the fusion
plasma community often named ``blobs''
\cite{krasheninnikov01,garcia06,nold10,dippolito11})
in an inhomogeneously magnetized e-p plasma are interchange unstable for a
range of parameters accessible in planned experiments and could lead to
crucial transport losses \cite{kendl17}. 

In the following it is demonstrated that replacing a fraction
of the positrons with ``impurity'' ions effectively reduces the interchange
propagation and transport of such e-p-i density filaments.
The relevance of filamentary transport lies in the self-propellation of such
elongated density perturbations down any magnetic field gradient, which does not
require a background temperature or density gradient for stimulating transport.
The initial perturbation may arise out of any plasma density or electric field
inhomogeneity.

It is here shown that both the effective mass dependent polarization inertia
(which is also active in a 2-d model) and the Boltzmann spin-up (which is a
3-d effect depending on Coulomb collisionality between the species) contribute
to the filament propagation reduction. 

In sec.~II the 3-dimensional full-F gyrofluid model used for the numerical
simulations is discussed, and in sec.~III it is argued why a delta-f model
(evolving only small fluctuations on a constant background plasma) is
inappropriate for describing interchange instability in e-p-i plasmas. The
(full-F) computational results are presented in sec.~IV, and conclusions are
given in sec.~V.  

\section{Full-F e-p-i gyrofluid model}

We analyse magnetized e-p-i plasmas by means of a nonlinear three-dimensional full-F
multi-species gyrofluid model, which is based on the 6-moment derivation of
Madsen \cite{madsen13} from a gyrokinetic model that evolves the full
distribution function $F({\bf   x}, {\bf v}, t)$, including a first order
finite Larmor radius (FLR) closure.

In the isothermal electrostatic limit \cite{wiesenberger14,kendl15} the full-F
3-d gyrofluid model consists of normalized continuity and momentum equations
for the gyrocenter densities $n_s$  and parallel velocities $v_s$ for
all species $s$, which here are given by electrons, positrons and ion with $s
\in (e, p, i)$.

\begin{eqnarray} 
\partial_t \hat n_s &=& {1 \over B} \left[\hat n_s, \phi_s \right] 
- {B \over n_s} \nabla_\pll  \left( { n_s v_s \over B} \right)
 + {\cal K}(h_s) \label{eq:den0}\\ 
\partial_t v_s &=& {1 \over B} \left[v_s, \phi_s \right]  
- {1 \over \mu_s} \nabla_\pll h_s
- {C_0 } J_\pll \nonumber \\ 
& & + \tau_s v_s \; {\cal K} (\hat n_s) + 2 \tau_s \; {\cal K} (v_{s}). 
\label{eq:vel0} 
\end{eqnarray}

Here $h_s \equiv  ( \phi_s + \tau_s {\hat n_s} )$ is abbreviated.
The (species specific) gyro-screened potentials $\phi_s =
\Gamma_{1s} \phi - (\mu_s/2B) (\nabla \phi)^2$ include both FLR and
ponderomotive effects.

The total parallel current $J_{||}$ is given by $J_{||} = \sum_s n_s Z_s e v_{s}$.
We have neglected triple nonlinear terms involving the parallel velocities, and
also electromagnetic fluctuations of the vector potential $A_\pll$.

The gyrocenter densities $n_s$ are normalized to a constant reference density
$n_0$, so that the magnitude of the plasma (electron) density $n_e \leftarrow n_e/n_0$ 
is of order one.
Eqs.~(\ref{eq:den0}, \ref{eq:vel0}) have been divided by the
gyrocenter densities $n_s$, so that logarithmic densities 
$\hat n_s \equiv \ln n_s$ are introduced as the evolving quantity to ensure
positivity, with both ${\hat n_s}$ and $n_s$ appearing in the equations. 
The potential is normalized to $T_e/e$, perpendicular length scales to the ion
drift scale $\rho =\sqrt{T_e m_i}/(eB)$, and time to $c_s/\rho$ with sound
speed  $c_s = \sqrt{T_e/m_i}$. $T_s$ and $m_s$ are the temperature and mass of
the species $s$, and $B_0$ is a reference magnetic field strength.
Parallel derivatives are further scaled as $\nabla_\pll \leftarrow
(L_\pll/L_{\perp}) \nabla_\pll$ with the connection length $L_\pll$, which for
toroidal geometry is given by $L_\pll = 2 \pi q R$ with inverse rotational
transform $q$ and major torus radius $R$.
The drift parameter $\delta = \rho_s / L_\perp$ is used to set the
perpendicular length scale $L_{\perp}$.  

The nonlinear quasi-neutral polarisation equation
\begin{equation}
\sum_s \left[ Z_s \; \Gamma_{1s} n_s + \bnabla \cdot \left( n_s {\mu_s \over
  Z_s  B^2} \bnabla \right) \phi \right] = 0.
\label{eq:pol0}
\end{equation}
determines the electrostatic potential $\phi$ for given gyrocenter densities $n_s$.

In the (2-d) model used in ref.~\cite{kendl17} for studying interchange
transport in pure e-p plasmas we had included Debye length effects into the
polarisation equation. This had restored the original ``Poisson'' term from the
electrostatic Poisson equation
$\sum_s Z_s N_s + \varepsilon \; \nabla_{\perp}^2 \phi = 0$,
in order to determine the (strongly damping) influence of Debye screening on
filament propagation. 
The Debye parameter  $\varepsilon = ( {\lambda / \rho})^2$ represents effects
of finite Debye length $\lambda = \sqrt{\epsilon_0 T_e/(e^2 N_{e0})}$ in
relation to the drift scale or Larmor radius $\rho_s=\sqrt{T_s m_s}/(eB)$
\cite{jenko02}.  
Here we specifically neglect these Debye effects and set $\varepsilon = 0$
(and thus can assume exact quasi-neutrality in the polarization equation), in
order to focus only on the influence of ion impurities on e-p filament
dynamics and to reduce the number of free parameters in the model. 

The particle densities $N_s$ are linked to the gyrocenter densities $n_s$ by
the relation
\begin{equation} 
N_s = \Gamma_{1s} n_s + \bnabla \cdot 
\left( n_s  {\mu_s \over Z_s  B^2} \bnabla_{\perp} \phi \right).
\label{eq:gyrocenter}
\end{equation}

The charge states are $Z_e=- 1$ for electrons and $Z_p=+1$ for
positrons. $Z_i$ depends on the impurity ion species and ionization degree,
but for the expected low temperatures of magnetized e-p laboratory plasma in
the range of a few eV we may assume predominantly singly ionized atoms or
molecules with $Z_i = +1$. Neutral impurities (and their ionization
and recombination processes) are here neglected. We also specifically neglect
electron-positron annihilation, which can for low e-p densities be expected to
occur on much longer time scales compared to the instability growth times
\cite{helander03}. 
In principle, annihilation (or positronium formation) rates could be easily
included as sink terms on the right hand side of eq.~(\ref{eq:den0}).

The gyro-averaging operator in Pad\'e approximation is defined by
$\Gamma_{1s} = (1 + (1/2) b_s)^{-1}$  with $b_s = \tau_s \mu_s \nabla_{\perp}^2$. 
The mass ratio is given by $\mu_s = m_s /(Z_s m_i)$, and the (constant)
temperature ratio by $\tau_s = T_s / (Z_s T_e)$. 
For electrons, thus $\tau_e = -1$, and for positrons we assume an equal
constant temperature so that $\tau_p = +1$. Ions are assumed to be cold with
$\tau_i \equiv 0$ so that also $b_i=0$.

Our model in principle can resolve all FLR effects, but in the following we will
neglect these also for electrons and positrons, and set $b_e = b_p = 0$.
Temperature dynamics and gradients could further influence the filament
propagation results \cite{held16}, but we here assume the e-p plasma to be
cool and isothermal. 

The 2-d advection terms are expressed through Poisson brackets
$[f,g] = (\partial_x f)(\partial_y g) - (\partial_y f)(\partial_x g)$ 
for local coordinates $x$ and $y$ perpendicular to ${\bf B}$.   
Normal and geodesic magnetic curvature enter the
compressional effect due to field inhomogeneity by  
${\cal K} = \kappa_y  \partial_y + \kappa_x  \partial_x$
where the curvature components in toroidal geometry
are functions of the poloidal angle $\theta$ mapped onto the parallel
coordinate $z$.  
For a circular torus $\kappa_y \equiv \kappa_0 \cos(z)$ and $\kappa_x \equiv
\kappa_0 \sin(z)$ when $z=0$ is defined at the outboard 
midplane. The toroidal magnetic field strength is assumed to vary only in parallel
direction as $B(z) = 1+a \cos(z)$ with inverse aspect ratio $a$. 

The collisionality parameter in the parallel velocity equation is given by
$C = 0.51 (\nu_e L_\perp / c_s)$. 
We note that the collisionality term in the corresponding (electromagnetic)
equation for the momentum given in ref.~\cite{kendl15} was written as
$C (J_\pll / n_s)$. However, the parameter $C \sim \nu_e$ also includes a
direct density proportionality in the electron/positron-ion collision
frequency $\nu_e \sim n_e$, which cancels the inverse density factor in the
collisionality term, so we here use a constant $C_0 \sim \nu_e(n_0$),
evaluated at a fixed reference density $n_0$. The weak density dependence
in the Coulomb logarithm is neglected. 

For numerical stability, a small perpendicular hyper-viscosity term $- \nu_4
\nabla_{\perp}^4 \hat n_s$ is added on the right hand side of
eq.~(\ref{eq:den0}), and in 3-d computations parallel viscous terms
$\nu_\pll \partial_z^2 \hat n_s$ and $\nu_\pll \partial_z^2 v_s$ are added to 
eqs.~(\ref{eq:den0}) and (\ref{eq:vel0}), respectively.
Boundary conditions in $y$ direction are periodic for 2-d simulations, and
quasi-periodic (shear-shifted flux tube) for 3-d simulations.
The further numerical methods are presented in ref.~\cite{kendl15}.

\section{Inadequacy of a delta-f model}

The common delta-f isothermal gyrofluid model \cite{scott05,kendl14} 
is regained by splitting $n_s = n_{s0} + \tilde n_s$ into a static constant
background density $n_{s0}$ and the perturbed density $\tilde n_s$. 
When $\tilde n_s / n_{s0} \ll 1$, the right hand sides of eqs.~(\ref{eq:den0}) and
(\ref{eq:vel0}) can be linearized by approximating 
$n_s \approx n_{s0}$ so that $\hat n_s \approx \hat n_{s0}
+ ({\tilde n}_s / n_{s0})$, and neglecting all nonlinear terms except the
Poisson bracket:
\begin{eqnarray} 
\partial_t \tilde n_s &=&  {1 \over B} [ \tilde n_s, \tilde \phi_s ]
- B \nabla_\pll  \left( { \tilde v_{s} \over B} \right) 
+ {\cal K}(\tilde h_s) \label{eq:denloc}\\ 
\partial_t {\tilde v_s}  &=& {1 \over B} [ \tilde v_s, \tilde \phi_s ]  
- {1 \over \mu_s} \nabla_\pll \tilde h_s + 2 \tau_s {\cal K} (\tilde v_{s}) -
{C_0 \over \mu_s} \tilde J_\pll \label{eq:velloc} 
\end{eqnarray}

The consistent delta-f polarisation equation in the high-$k$ limit
is $\sum_s  a_s [ \Gamma_{1s} \tilde n_s + (1/\tau_s) (\Gamma_{0s} -1)
  \tilde \phi ]  =  0$ with  $\Gamma_{0s} = (1 + b_s)^{-1}$.
Linearisation of the low-$k$ eq.~(\ref{eq:pol0}) does not include
gyro-screening on the potential and gives 
$\sum_s  a_s \Gamma_{1s} \tilde n_s  = (\sum_s a_s \mu_s) \nabla_{\perp}^2 \tilde \phi$.  
The velocities and current are coupled in the electrostatic limit by
$\tilde J_{||}  =  \sum_s a_s \tilde v_{s}$.
The parameter $a_s = Z_s n_{s0}/ n_{e0}$ denotes the ratio of species
reference densities $n_{s0}$ to $n_{e0}$.

These delta-f gyrofluid equations are a good approximation to the full-F model for
example in core and mid-pedestal e-i fusion plasma turbulence simulations
\cite{scott10}, where density fluctuations indeed are usually much smaller
than the average background plasma density.

The applicability of the delta-f multi-species model, which assumes a quasi
infinite background density for all of the species, however specifically
fails, when one of the species has a much smaller or vanishing density
compared to the others.

Then the term
${\cal K}({\tilde h_s}) = {\cal K}( {\tilde \phi}_s + \tau_s {\tilde n_s} )$
inconsistently would generate by
$\partial_t {\tilde n_s} \sim {\cal K}({\tilde   h_s})$
new density out of any appearing inhomogenous potential fluctuation ${\tilde \phi}_s$
even if the initial species density (fluctuation) $\tilde n_s$ was zero.

This artefact is not present in the full-F model, which is evident when we do
not write eq.~(\ref{eq:den0}) in terms of the logarithmic density $\hat n_s$,
but originally as
\begin{equation} 
\partial_t n_s =  {1 \over B} \left[n_s, \phi_s \right] 
- {B} \nabla_\pll  \left( { n_s v_s \over B} \right)
 + n_s \; {\cal K}(h_s). 
\end{equation}
Here the interchange curvature term $ n_s \; {\cal K}(h_s)$ can drive changes
in density only in proportion to the locally present species density $n_s({\bf x},t)$.

\section{E-P-I filament propagation}

Pressure perturbations in magnetized plasmas experience interchange forcing
due to an inhomogeneity (gradient and curvature) of the magnetic field, which
leads to a ``radial'' propagation across the magnetic field.
Perturbations in toroidal plasmas are mostly flute-like and strongly
elongated along the magnetic field direction, and appear as plasma filaments.
In the fusion plasma literature, filamentary pressure perturbations at the
plasma edge with positive amplitudes are commonly named ``blobs'', or
``holes'' for negative amplitudes.

The basic gradient and curvature drift dynamics, which differ in sign between
positive and negative plasma species by their charges, results in an $E \times B$
drift which radially advects the perturbation. Initially symmetric
(e.g. Gaussian) shapes of the perturbation across the field then develop into
mushroom shaped plume structures. These effects are already present in 2-d
(perpendicular to the magnetic field direction) fluid models.

Along the magnetic field direction the dynamics is usually more wave-like
(compared to the fluid-like advection across the field), and pressure
perturbations can induce sound waves or Alfv\`en waves (which are however
neglected in the present work, with $\beta =0$). Collisional coupling between
the species can lead to deviations from an adiabatic response on perturbations.

When the initial perturbation filaments are extended with constant amplitude
everywhere along the magnetic field ($z$) direction, the dynamics again
becomes quasi-2-dimensional (except for effects of magnetic shear).
Here we rather consider perturbations that are also initially localized in the
parallel direction with some parallel width $\Delta z$ in the maximum
ballooning region (which here defines $z=0$). Then the perturbation will
experience spreading along the field direction by pressure driven expansion.

The sound speeds of electrons and positrons are (for equal temperature)
identical due to the same mass, but the sound speeds differ substantially
between electrons and the much more massive ions. The more rapid excursion of
electrons from (initially neutral) pressure perturbations together with ion
inertia leads to a positive charging of the perturbation, which again slows
down the electrons into an ambipolar parallel diffusion. In e-i plasmas this
arising potential perturbation leads to a vortical $E \times B$ drift around the
perturbation, which spins the blob into an eddie and effectively slows down
the radial interchange drive of the whole filament. This effect, named 
``Boltzmann spinning'' in the fusion plasma literature \cite{angus12}, is absent
in pure mass-symmetric pair plasmas.

In the following, effects of the presence of some fraction of ion impurities in
an e-p pair plasma on interchange driven filament transport will be studied.
Boltzmann spinning of localized perturbations may be expected to slow
the e-p blobs depending on ion concentration. This effect will be addressed
with 3-d simulations.

\subsection{Inertial mass effect through polarization}

But already in a simplified 2-d setup another species mixture effect on blob
propagation can be expected by changes of the effective mass of the plasma,
which enters into polarisation dynamics mediated by eq.~(\ref{eq:pol0}).

Linearisation of the polarisation equation (without FLR effects) gives
\begin{equation}
\sum_s  a_s \tilde n_s  = (\sum_s a_s \mu_s) \nabla_{\perp}^2 \tilde \phi
\equiv \bar \mu \; \Omega.
\end{equation}

The development of $E \times B$ vorticity $\Omega = \nabla_{\perp}^2 \tilde \phi$
out off density perturbations is thus mediated by an effective mass $\bar \mu
= \sum_s a_s \mu_s$. Here $a_s = Z_s n_{s0}/ n_{e0}$ and $\mu_s = m_s / (m_e
Z_s)$, when the electron mass $m_e$ is used as a reference. For a pure e-p
pair plasma (with $a_i=0$), $a_e = \mu_e = -1$ and $a_p = \mu_p = +1$, so that
$\bar \mu = 2$.

The effective mass in an e-p-i plasma is given by 
$\bar \mu = a_e \mu_e + a_p \mu_p + a_i \mu_i = (a_e - a_p) \mu_e + a_i \mu_i$.
When the electron density is kept constant and a fraction of positrons
in an e-p plasma is replaced by ions, then $a_e = - 1$ and $a_p = 1 - a_i$, so that 
$\bar \mu = 2 + (\mu_i -1) a_i \approx 2 + \mu_i a_i$.

The effects of variations in the effective mass on interchange driven filaments
and turbulence in the edge and scrape-off layer of tokamak fusion plasmas has
been recently investigated for the similar ion masses in hydrogen isotope mixtures
\cite{meyer16,meyer17,meyer17B}, where relevant changes have been found, so
that even stronger effects can be expected for the e-p-i system with large
mass differences between the positive species. Significant changes should
occur at least for ion density ratios $a_i > \mu_i^{-1}$. 

In the following numerical examples we assume hydrogen ions as the impurity
species, so that $\mu_i = m_i / (m_e Z_i) \approx 1836$. When the impurities
are generated by e-p plasma-wall interactions or by rest gas contamination in
an imperfect vacuum chamber, the ion masses can be larger, depending on the
present atomic or molecular species. For thermal ionization the charge state $Z_i$
of impurity ions will likely be single, but ionization by annihilation photons
could lead to stronger degrees of ionization in impurity species.

For simplicity we here thus only consider hydrogen ions (protons) and keep in mind
that heavier species would more enhance the reported mass effects. The ion
contribution to mass inertia is of order unity and larger when $a_i \mu_i
>1$, or when $a_H > 1/1836 \approx 5 \cdot 10^{-4}$. For comparison, for singly
charged iron impurities (from the chamber wall) with $\mu_{Fe} \approx 56 \;
\mu_H$, the critical concentration, above which inertial mass effects become
relevant, would be around $a_{Fe} \sim 10^{-5}$.

We first investigate the inertial mass effect through $\bar \mu$ in 2-d
simulations. For this we numerically evaluate eqs.~(\ref{eq:den0}) and
(\ref{eq:pol0}) for $v_s = 0$ at the location $z=0$, where the normal
curvature is maximum and the geodesic curvature contribution vanishes.
As mentioned above, this corresponds to a case of highly elongated filaments.

A Gaussian initial density perturbation with perpendicular width $\sigma = 4 \rho$ and
amplitude $\Delta n_{e} = 0.5$ is set on an otherwise homogeneous density
background with $n_{e0} = 1.0$. The magnetic curvature is set to $\kappa_0 = 0.01$.
The computational domain is $L_x \times L_y = 64 \rho \times 32 \rho$ on a
rectangular numerical grid with $n_x \times n_y = 128 \times 64$.
Higher grid resolution leads to nicer resolved pictures of the blobs, but does
not change the results significantly.

A fraction $a_i$ of positron density is replaced by (hydrogen) ions, and is varied
between $a_i=0$ for a pure e-p pair plasma, up to $a_i =1$ for a pure e-i plasma.

\begin{figure} 
\includegraphics[width=8.0cm]{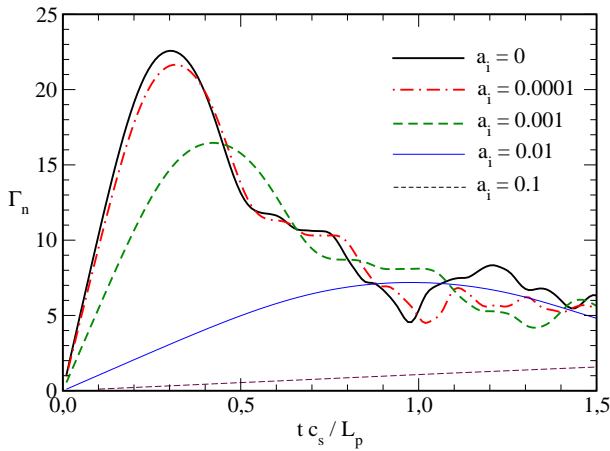}
\caption{2-d numerical results: density transport $\Gamma_n(t)$ by interchange
  driven $E \times B$ advection of electron-positron plasma blobs for various
  values of an ion impurity fraction $a_i$.}   
\label{fig-2dgam}
\end{figure}

The average interchange transport by radial blob propagation is determined by
$\Gamma_n = \langle n_e v_x \rangle_{x,y}$ with $v_x = \partial_y \phi$.
The transport as a function of normalized time is shown in
Fig.~\ref{fig-2dgam}: it increases to a maximum as long as the radial blob
propagation velocity accelerates, and then drops again to low
levels. Nonlinear breakup of the blob leads to a more unsymmetric decay phase.
We observe that the maximum transport (as well as the maximum blob velocity)
is strongly reduced with increasing ion fraction by a factor $1 / \sqrt{\bar
  \mu}$.

\begin{figure} 
\includegraphics[width=8.0cm]{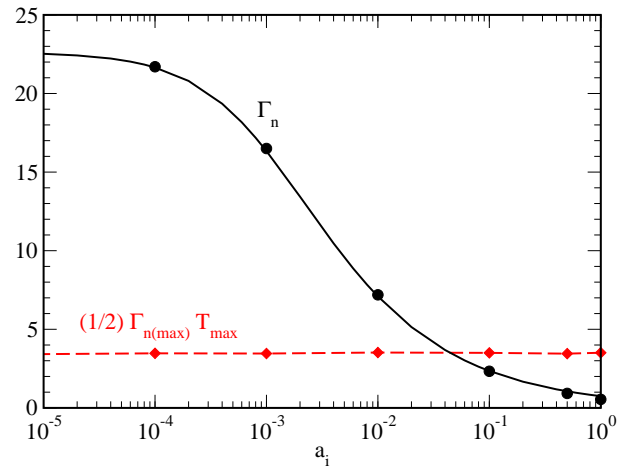}
\caption{Maximum 2-d transport (black dots) and average integrated transport (red
  diamonds) of electron positron blobs as a function of the (hydrogen) ion
  impurity fraction $a_i$.}  
\label{fig-2dai}
\end{figure}

The time scale $T_{max}$ for acceleration, until the maximum velocity and transport
level are reached, on the other hand grows with increasing ion fraction: the
interchange growth rate and propagation velocity are shifted from electron to
ion time scales by the same factor $1 / \sqrt{\bar \mu}$.
The total integrated transport, which we here (because of the nearly symmetric
shape of $\Gamma_n(t)$) approximate as $\int \mbox{d} t \; \Gamma_n(t) \approx (1/2)
\Gamma_{max} T_{max}$, is therefore largely independent of the ion impurity
fraction, but is only spread over different time scales.
Both the maximum transport and the approximate integrated transport are shown
as a function of the ion fraction in Fig.~\ref{fig-2dai}. The black dots are the
numerical values of the maxima from the simulations shown in
Fig.~\ref{fig-2dgam}, and the black line is the analytical function
$\Gamma_n(a_i) = \Gamma_n(0) /\sqrt{\bar \mu}$, with $\bar \mu = 2+1836 \; a_i$.

\subsection{Boltzmann spinning effect}

\begin{figure}
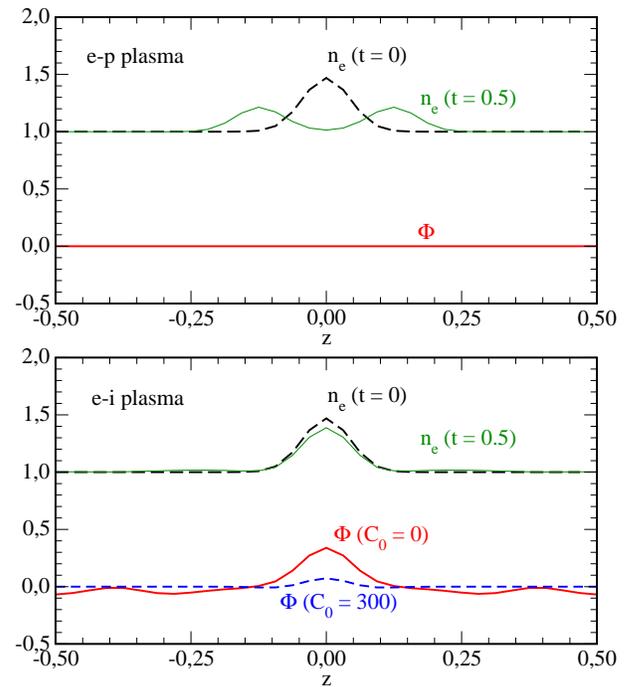
 
\includegraphics[width=8.0cm]{fig3a-3das1.0.epsi}
\includegraphics[width=8.0cm]{fig3b-3das0.0.epsi}
\caption{Parallel spreading of $z$-localized 3-d filaments. Top: In a mass
  symmetric e-p pair plasma the density propagates with the
  electron/positron sound velocity; no electric potential develops. Bottom: In
an e-i plasma the electrons pull outward but are restrained by a potential
$\phi(z)$; the filament remains more coherent, depending on collisionality $C_0$.}  
\label{fig-3dpara}
\end{figure}

The inertial polarization mass effect, which effectively scales the blob
propagation time, is also still present for 3-d simulations of elongated filaments.
For a finite initial filament extension $\Delta z$ along the field line, the
time scale of radial advection then competes with the time scales of parallel
spreading and charging.

\begin{figure}
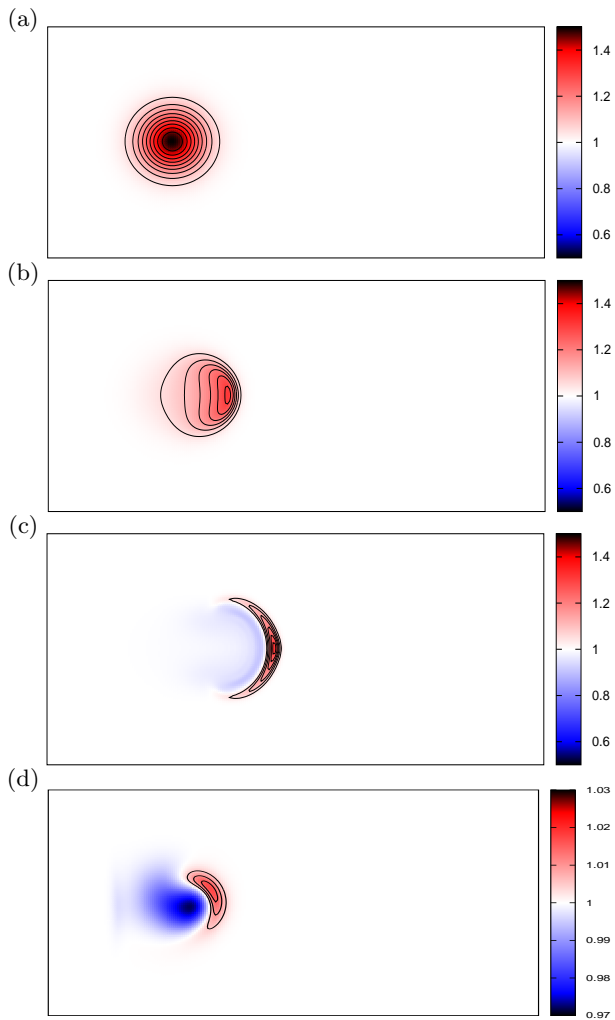
 
(a) \includegraphics[height=7.5cm, width=3.1cm, angle=270]{fig4a-as1-t00.epsi}\\
(b) \includegraphics[height=7.5cm, width=3.1cm, angle=270]{fig4b-as1-t50.epsi}\\
(c) \includegraphics[height=7.5cm, width=3.1cm, angle=270]{fig4c-as1-t100.epsi}\\
(d) \includegraphics[height=7.5cm, width=3.1cm, angle=270]{fig4d-as0.99-t100.epsi}
\caption{Evolution of a density blob (2-d $x$-$y$ cross sections of 3-d
  simulations) in an inhomogeneous magnetic field. From top to bottom: (a) initial
perturbation at $t=0$; (b) e-p blob at $t=50$; (c) e-p blob at $t=100$; (d):
e-p blob with 1\% ion fraction, showing Boltmann spin-up.}   
\label{fig-evolve}
\end{figure}
In 3-d the collisionality $C_0$, the filament extension $\Delta z$, and the
parallel-to-perpendicular scale ratio $\hat \epsilon = (qR/L_{\perp})^2$ enter
as additional parameters and control the non-adiabatic electron response. 

The difference in electric potential generation by parallel evolution for pure
e-p compared to e-i plasmas is shown in Fig.~\ref{fig-3dpara}:
A $z$-localized e-p blob (top) propagates its perturbation in both directions
along the field line with exactly the electron/positron sound velocity, but
the electric potential $\phi$ remains zero.
The e-i blob (bottom) on the other hand remains after the same time ($t=0.5$
in $c_s/L_{\perp}$ normalized units) more coherent in the parallel direction,
but develops an electric potential, which follows a Boltzmann relation
$\phi \sim n_e$ for the adiabatic case ($C_0 =0$) and is weaker for a strongly
collisional case ($C_0=300$).

As the blob is not only localized in $z$-direction but also is initialized
with a Gaussian bell shape in perpendicular $x$-$y$ direction, the development
of an aligned electric potential leads to the onset of $E \times B$ advection
azimuthally around the perturbation with the drift velocity $v_{E \times B} =
(1/B^2) {\bf B} \times \bnabla \phi$. This rotates the blob differentially
around its axis, the ``Boltzmann spinning''. 

The influence of Boltzmann spinning on e-p blobs with ion impurities is shown
in Fig.~\ref{fig-evolve} as 2-d $x$-$y$ cross sections (at $z=0$) of 3-d
simulations for various times.
The cross section shows the computational region of $64 \rho \times 32
\rho$ like in the other simulations above, and the blob is again initially
localized with $\sigma = 4 \rho$, now with parallel width $\Delta z = 1/8$.  
On the top, picture (a) shows the initial density $n_e({\bf x}, t=0)$
with the same Gaussian 
perturbation for both e-p and e-i blobs. Picture (b) and (c) show the
evolution of a pure e-p blob at the times $t=50$ and $t=100$, respectively.
The radial propagation velocity and the associated outward density transport
are maximal at around $t=50$.
Picture (d) shows at $t=100$ the onset of density spin-up by $E \times B$ drift
advection in an e-p blob with 1~\% ion impurity fraction ($a_i = 0.01$), which
effectively suppresses the outward interchange driven propagation of the
filament. Note that the density color scale has been adapted in (d), as the blob
also looses amplitude at the shown location at $z=0$ due to parallel spreading.

\begin{figure} 
\includegraphics[width=8.0cm]{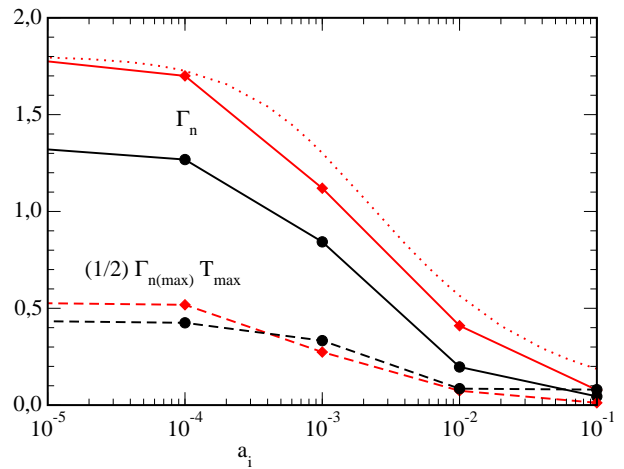}
\caption{Maximum interchange transport (straight line) and average integrated
  transport (dashed line) of 3-d electron positron filaments as a function of
  the (hydrogen) ion impurity fraction $a_i$, for collisionalities $C_0=30$
  (black / circles) and $C_0=300$ (red / diamonds).}  
\label{fig-3dai}
\end{figure}

The combined effect of mass inertia and Boltzmann spinning on filamentary e-p
transport is now computed for varying ion impurity densities. 
In Fig.~\ref{fig-3dai} the maximum transport (averaged
over the parallel coordinate) is again shown as a function of the ion fraction
$a_i$. Black lines and circle symbols denote the simulation results for
$C_0 = 30$, while the red lines and diamond symbols denote the results for $C_0=300$. 
Further simulation results (not shown here) for a completely adiabatic
response with $C_0=0$ are nearly identical (slightly smaller) compared to the
results for $C_0=30$. This range approximately covers values that may be
expected for low-temperature e-p laboratory plasmas of a few eV. 

The dotted red line shows the analytical estimate for the inertial mass effect
on $\Gamma_n(a_i) = \Gamma_n(0) /\sqrt{2+1836 \; a_i}$ for $C_0=300$.
While in the 2-d case the simulation results were nearly exactly lying on the
analytical graph, we here see a sytematically lower transport level, although
the transport still approximately follows the overall analytical trend of inertia.
The further reduction is a combination of density decrease by parallel
spreading and of a suppression of radial filament propagation by Boltzmann
spinning.
While the 2-d integrated transport was independent of the ion impurity
fraction, we here observe a significant reduction of the values for
$(1/2) \Gamma_{n(max)} T_{max}$ with $a_i$ (depicted by the dashed black and red lines
  connecting the simulation values) by the 3-d Boltzmann spinning effect.
The strongest change of filament transport by $a_i$ in Fig.~\ref{fig-3dai}
still occurs for values around $\mu_i a_i \sim 1$, which for the presently
assumed hydrogen ions is for $a_i \sim 1/1836 \approx 0.5 \cdot 10^{-3}$.
For more massive impurities the ion effect on e-p interchange transport would
accordingly occur already for lower density fractions.

\section{Conclusions and outlook}

To summarize, we have presented the first computations of interchange
transport in inhomogeneously magnetized e-p plasmas with impurity ions.
The reduction of transport with increasing ion fraction $a_i$ roughly follows
the inertial mass scaling, and is additionally reduced by Boltzmann spinning
which depends on parallel localization of the filament and on the dissipative
parallel coupling between leptons and ions. 

Is this effect, after all, in any way relevant? Can, for example, a
significant impurity density be expected in planned e-p confinement
experiments? The parameters of future experiments \cite{pedersen12}, like
achievable e-p densities, temperatures, or radial profiles have large
uncertainties. Any reliable predictive theoretical modelling of confinement
properties and expectable modes and instabilities is thus not honestly possible.
Theory can for now only stake out likely effects and trends. 

The higher edge temperatures and more energetic edge localized transport
events in magnetized fusion plasmas lead to sputtering and erosion of the
plasma-facing wall components, which may enter the confined plasma region as
impurity ions. In low-temperature e-p experiments the impurity content may be
much lower, but also depends on the purity of the initial vacuum. For iron
impurity ions from the vacuum chamber, the critical concentration where
interchange mass effects would become noticeable is around $a_i \sim 10^{-5}$,
which is not completely unrealistic. But only the first real experiments will be
able to clarify the e-p plasma purity.

On the other hand, the impurity concentration could also be set on purpose to
probe the e-p to e-i transition by injecting for example hydrogen ions into a
confined e-p plasma. Such dedicated e-p-i experiments would be able to test and
validate our theories and models of plasma physics, which would be of general
value for other areas like magnetic confinement fusion research.

However, in low density e-p plasmas the mass effect on interchange driving
will not appear alone but in context with the Debye screening studied in
ref.~\cite{kendl17}. There we had derived the interchange growth rate (and
accordingly the radial propagation velocity and associated transport) to be
proportional to $\gamma \sim 1/\sqrt{\hat \mu + \epsilon}$, where
$\hat \mu = 2 + a_i \mu_i$ was fixedly set to 2 for the pure e-p plasma.
The values for the Debye parameter $\epsilon$ have been estimated to be in the
range of 50-300 for planned experiments. This implies that any ion impurity
concentration effect will only become inertially relevant for $a_i \mu_i$ in a
similar order of magnitude as $\epsilon$, or above. 
For a value of $\epsilon = 200$ and hydrogen ions, this would require
a concentration of around 10~\%, which appears to be unrealistically large for
chance wall or rest gas impurities to be of any relevance. If the ion mass
effect and any e-p to e-i physics transition should be tested on purpose, then
a larger e-p plasma density in the order of the Brillouin density would be
required in the experiments to overcome the Debye damping.
So it can be concluded that an ion mass effect on interchange transport for
vanilla operating conditions is likely to be subdominant.

However, we have so far ignored an additional possible ion
impurity mechanism in magnetized e-p plasmas: the presence of ions is expected
to be able to trigger the onset of resistive drift wave or drift-Alfv\'en wave
instabilities and associated turbulence in e-p plasmas in the presence of a
background density gradient. The turbulent transport resulting from e-p-i drift wave
turbulence may still turn out to be detrimental for magnetic e-p confinement,
if it is not also effectively damped by Debye shielding.
The computational investigation of fully developed e-p-i drift wave turbulence
is however rather expensive because of the high required resolution
to resolve the disparate electron/positron and ion drift scales appropriately.
In particular the necessity to use full-F models, which are computationally
also much more demanding than delta-f models, presently slows down the acquisition
of results. These will therefore have to be reported in a future work.

\end{document}